\baselineskip=20pt
\magnification=\magstep1
\centerline{A Precessing Disc in OJ287?}
\bigskip
\centerline{J. I. Katz}
\centerline{Department of Physics and McDonnell Center for the Space
Sciences}
\centerline{Washington University, St. Louis, Mo. 63130}
\centerline{katz@wuphys.wustl.edu}
\bigskip
\centerline{Abstract}
\medskip
Sillanp\"a\"a, {\it et al.} (1996) have demonstrated that the AGN OJ287 has
intensity peaks which recur with a period of about 12 years.  I suggest that
this is the result of the sweeping of a precessing relativistic beam across
our line of sight.  In analogy to Her~X-1 and SS433, precession is
attributed to the torque exerted by a companion mass on an accretion disc.
Secondary maxima observed 1.2 years after two of these peaks may be evidence
of nodding motion.
\vfil
\noindent
Subject headings: Galaxies: BL Lacertae Objects: OJ287 --- Accretion,
Accretion Discs \par
\eject
\centerline{1. Introduction}
\smallskip
Sillanp\"a\"a, {\it et al.} (1996) reported, in confirmation of the
prediction of Sillanp\"a\"a, {\it et al.} (1988), that the AGN OJ287 has
outbursts in visible light with an observed period of about 11.65 y (8.9 y
in its rest frame at its cosmological redshift of 0.306).  This
extraordinary result marks the first confirmed periodicity of any
extragalactic object, other than variable stars observed in nearby
galaxies.

Several mechanisms, generally assuming a binary supermassive black hole,
have been proposed which may explain this periodicity.  Begelman, Blandford
\& Rees (1980) (henceforth BBR) suggested that accretion discs and jets
surrounding a supermassive black hole may undergo geodetic precession in the
gravitational field of a binary companion black hole.  This process would
produce uniform precession which would modulate a jet's observed Doppler
shift and its observed intensity.  Typical estimates of the geodetic
precession period are several hundred years, much longer than observed in
OJ287.

Sillanp\"a\"a, {\it et al.} (1988) suggested that the observed period might
be the binary period itself, with the companion (another massive black
hole, or conceivably a star) disrupting an accretion disc and modulating its
accretion rate.  These authors suggested a strongly eccentric and rapidly
relaxing orbit, but such a short period orbit would probably have
circularized (BBR).  An inclined (to the accretion disc) circular orbit
might be more plausible, with the companion disrupting the disc and
stimulating accretion on each passage through it, implying an orbital period
of twice the observed period.  However, it is unclear how a local disruption
or tidal perturbation in the outer portions of an accretion disc could
produce a brief ($\sim 0.01$ of the orbital period) surge of accretion at
its center.  Further, a massive companion in a inclined orbit would make the
disc's axis precess about the total angular momentum axis and would deplete
the disc of material at the radius of the companion's orbit; a sudden and
disruptive plunge of the companion through the disc is likely only if the
companion's orbit is very eccentric.

The observed narrow spikes of intensity suggest a relativistic beam sweeping
close to or across our line of sight, in accordance with models of OJ287 and
similar objects which hypothesize such a beam directed nearly towards the
observer.  This paper proposes a model which attributes this beam
geometry to the precession of an accretion disc driven by the gravitational
torque of a companion mass.  In \S2 I review the properties of driven
accretion discs, and compare them qualitatively to observations of OJ287.
In \S3 I attempt to constrain the parameters of OJ287.  This requires
consideration of the possible evidence for nodding motions in OJ287 and its
implications.  \S4 contains a brief summary discussion.
\medskip
\centerline{2. Driven Precessing Accretion Discs}
\smallskip
An accretion disc inclined to the orbital plane of a binary system will
precess at the rate
$$\Omega_0 = - {3 \over 4} {G m_2 \over a} \left({a_d \over a}\right)^2
{\cos\theta_0 \over (G m_1 a_d)^{1/2}}, \eqno(1)$$
where $m_1$ is the mass of the accreting object, $m_2$ the mass of its
companion, $a$ their separation (assuming a circular binary orbit), $a_d$
the disc radius and $\theta_0$ the inclination of the disc to the orbital
plane.  The theoretical driven precession frequency $\Omega_0$ is
distinguished from the observed precession frequency $\omega_{pre}$;
$\Omega_0$ is a measure of the torques which drive the nodding motion, even
if the actual precession has other contributions and occurs at a different
rate (as is the case in SS433, for which $\Omega_0 / \omega_{pre} \approx
2.1$).  The physical mechanism is the same as that which drives the
recession of the nodes of the Moon's orbit, and was applied to the accretion
disc in Her~X-1 by Katz (1973) and to that in SS433 by Katz (1980).  This
Newtonian driven precession is, in general, much faster than geodetic
precession.

Observations of Her~X-1 and SS433 provide sensitive measures of the behavior
of their precessing discs.  Her~X-1 is eclipsed by sharp-edged disc
structures.  Accurately observed Doppler shifts in SS433 permit accurate
determinations of the orientation of its jets and disc.  Long time series
have permitted detailed investigation of these laboratories for the study of
accretion disc dynamics, and the results may be compared to observations of
OJ287:
\item{1.} The $Q$ of the precession, considered as an oscillator, is about
39 in Her~X-1 (Baykal, {\it et al.} 1993) and about 75 in SS433 (Baykal,
Anderson \& Margon 1993).  This is comparable to the $Q \sim 25$ implied by
the scatter in intervals between peaks of OJ287 reported by Sillanp\"a\"a,
{\it et al.} (1988), and contrasts to orbital periods or geodetic precession
periods, which should either be good clocks with very high $Q$ or show
monotonically decreasing periods if dissipative processes shrink the orbit.
\item{2.} In addition to its mean precession there is an oscillation
(``nodding'') in the disc's orientation with frequency $2\omega_{orb} -
2\omega_{pre}$ ({\it{N.B.:}} the orbital frequency $\omega_{orb}$ is positive
by convention and $\omega_{pre}$ is negative).  In Her~X-1 nodding produces
preferred orbital phases for the X-ray source's emergence from eclipse by
the accretion disc and complex pre-eclipse dip behavior; in SS433 it
produces a 6-day period in the Doppler shifts (Katz, {\it et al.} 1982,
henceforth KAMG; Levine and Jernigan 1982).  In OJ287 nodding may explain
the secondary peaks observed 1.2 years after the main peaks in 1971 and
1983; only the driven precessing disc model naturally explains them.
Flickering in the intensity of OJ287 makes it difficult to identify the
nodding motion except near the peak.
The secondary peaks are not explicable if the 12 year period is attributed
either to the orbital period or to geodetic precession.
\medskip
\centerline{3. Parameters of OJ287}
\smallskip
In order to apply the precessing disc model to OJ287 we should estimate the
critical parameters $i$ and $\theta_0$, where $i$ is the inclination of the
orbital angular momentum axis to the direction to the observer.
Unfortunately, in contrast to SS433 no quantitative kinematic information
is available for OJ287.  We can, however, make some estimates, noting that
the beam-width of radiation emitted by a relativistically moving object is
$\sim 1/\gamma$, where $\gamma$ is its Lorentz factor of bulk motion (it is
assumed to radiate isotropically in a frame which has this Lorentz factor
with respect to the observer's frame, although this is surely an
oversimplified description of a cloud of relativistic electrons directed
approximately in our direction).  In order that the observer be within the
path of the beam we must have
$$\vert i - \theta_0 \vert {\buildrel < \over \sim} 1/\gamma + \theta_n,
\eqno(2)$$
where $\theta_n$ is the amplitude of the nodding modulation of the
precession angle.  

In order to produce an intensity peak whose half-width (Sillanp\"a\"a, {\it
et al.} 1996) is a fraction $f \sim 0.01$ of the precession period requires
a beam width $\sim 2 \pi f \sin\theta_0$, where a factor-of-two uncertainty
is introduced by the fact (KAMG) that nodding multiplies the rate of
precession around its mean path by a factor varying (with the nodding
frequency) from 0 to 2, as well as introducing periodic oscillations in
$\theta$ about its mean value $\theta_0$.  This factor, as well as the
comparative intensities of the primary and secondary peaks (and their exact
separation) depend on the relative phases of the nodding and orbital motion,
and will not repeat from cycle to cycle unless the motions happen to be
commensurate.  Then
$$\gamma \sim (2 \pi f \sin\theta_0)^{-1} \sim 50, \eqno(3)$$
(where $\theta_0 = 20^\circ$ is used in analogy to SS433, the only system
in which $\theta_0$ is known).  This is larger than the $\gamma \sim 10$
typically inferred from radio measurements of superluminal expansion in AGN,
but is rather uncertain; source selection on the basis of visible intensity
measurements may introduce a bias towards large Lorentz factors so that the
very luminous OJ287 may be characterized by larger $\gamma$ than most AGN.
In addition, visible emission may result from more relativistic motion than
that characterizing the more extended regions (and less energetic particles)
producing radio emission.

If the secondary maximum in intensity observed 1.2 y after the 1971 and 1983
peaks is attributed to nodding then the orbital period is determined as well
as the precessional period, and it becomes possible to determine other
parameters of OJ287.  The nodding amplitude is (KAMG)
$$\theta_n = {\vert \Omega_0 \vert \tan\theta_0 \over 2(\omega_{orb} -
\omega_{pre})}. \eqno(4)$$
If $\Omega_0 = \omega_{pre}$ (likely in a supermassive black hole binary,
for which the mechanisms leading to $\Omega_0 \ne \omega_{pre}$ in mass
transfer binary stars are inapplicable) then the observed parameters,
attributing the secondary intensity peak to nodding motion, imply $\theta_n
\approx 0.11 \tan\theta_0$.  The precession amplitude $\theta_0$ is unknown,
but again assuming $\theta_0 = 20^\circ$ as in SS433 suggests $\theta_n \sim
0.04$.  If the secondary peak is not attributable to nodding then
$\omega_{orb}$ is unknown; analogy to Her~X-1 and SS433 would then suggest
an orbital period in the range 0.5--1 y and $\theta_n \approx (0.02$--$0.04)
\tan\theta_0 \sim 0.007$--$0.014$.

Combining Eq. (3) and our estimated $\theta_n$ implies that $i$ and
$\theta_0$ must be equal to within roughly $\pm 3^\circ$ (the actual
condition is uncertain because $\theta_0$ is unknown).  Approximate equality
between these two angles is required in any precession model which is to
explain the narrowness of the periodic intensity peak (a broader peak would
lead to a lower estimated $\gamma$ and a less stringent constraint on $\vert
i - \theta_0 \vert$), and does not depend on the precession mechanism.  The
permitted angular range, roughly a third of $\theta_0$, is broad enough to
be plausible, in part because the presence of nodding makes Eq.~(2) less
stringent than it would otherwise be; nodding increases the width of the
swath about the mean precession path over which the beam is swept.
Intensity selection introduces a strong bias in favor of detection of
sources whose beams are, at least occasionally, directed towards the observer.

Nodding produces two periodic terms in the jet's Doppler shift.  Their
amplitudes (in the redshift $z \equiv \Delta \lambda / \lambda_0$) are
(KAMG)
$$\eqalign{A(2\omega_{orb} - 2\omega_{pre})&= {\gamma v \Omega_0 \sin\theta_0
\tan\theta_0 \cos i \over 2 (\omega_{orb} - \omega_{pre})}\cr
A(2\omega_{orb} - \omega_{pre})&= {\gamma v \Omega_0 \sin\theta_0 \sin i
\over 2 (\omega_{orb} - \omega_{pre})}.\cr}\eqno(5)$$
The ratio of these amplitudes is $\tan\theta_0 \cot i$.  In SS433 the second
of these amplitudes is dominant because $i$ is nearly 90$^\circ$ but
$\theta_0$ is small, but in OJ287, with $i \approx \theta_0$ expected, both
amplitudes should be comparable.  However, in OJ287 the observed signal is
an intensity which is proportional to $(1 + z)^{-(2 + \alpha)}$, where
$\alpha$ is the spectral index (typically $\alpha \approx 0.5$), and hence
is narrowly peaked around the minimum in $1 + z$ when the beam points nearly
directly towards the observer, in contrast to SS433 in which the periodic
variations in Doppler shift are directly observed throughout the
precessional cycle.  The implied orbital period in the source frame is 2.1 y
or 2.3 y, depending on which of the two periods in Eq.~(5) is identified
with the observed 1.2 year interval.

The separation of the two black holes is 
$$a \approx 1.1 \times 10^{16} \left({m_1 + m_2 \over
10^8\,M_\odot}\right)^{1/3}\ {\rm cm}.\eqno(6)$$
In Eq.~(6) the lower of the two possible nodding frequencies was assumed,
although the difference between them is less than the error in determination
of the separation between the peaks because of flickering in the intensity.
The lifetime to gravitational radiation (BBR) is then
$$t_{gr} \approx 4 \times 10^5 \left({m_1 \over 10^8\,M_\odot}\right)^{-5/3}
{(1 + \mu)^{4/3} \over \mu}\ {\rm y}, \eqno(7)$$
where $\mu \equiv m_2/m_1$.

Comparison of the precessional to the orbital period permits, using
Eq.~(1), estimation of the size of the precessing disc.  It approximately
fills the Roche lobe, and is about the maximum size permitted for a stably
orbiting ring (Bahcall, {\it et al.} 1974); for such a large ring Eq. (1) is
not accurate and the ring may be smaller than indicated and have more complex
motion.  OJ287 differs from Her X-1 and SS433, whose much smaller precessing
rings fit well within their Roche lobes.  This is not surprising, for in a
mass transfer binary the accretion disc is fed by matter from the companion
star through its wind or Roche lobe overflow with comparatively little
specific angular momentum.  In an AGN accretion occurs from an extended
region outside the companion's orbit, fed by distant stellar disruptions and
mass loss with large specific angular momentum with respect to the accreting
mass.  The secondary mass may then be less effective in disrupting the outer
parts of the disc than in a mass transfer binary, especially if $\mu \ll 1$.
\medskip
\centerline{4. Discussion}
\smallskip
The observation of periodicity in OJ287, the AGN with perhaps the best long
time series of brightness data, suggests that this phenomenon is common.
Similarly, the best observed eclipsing accretion disc (that of Her X-1) and
the only observed subrelativistic jet source (SS433) show precession,
suggesting that precession is frequent, if not universal, in discs: wherever
the data are good enough to reveal precession, it is found.

Can this be an accident?  In binary X-ray sources it has been suggested
(Katz 1973) that precession may be self-excited as the accretion disc
shadows the secondary star and influences the geometry and angular
momentum of radiation-driven mass transfer; precession would then be
expected whenever the accreting object is luminous.  This mechanism is
probably inapplicable to binaries composed of two black holes.  In an AGN
there is no {\it a priori} reason to expect the angular momentum of accreted
gas, whose origin is stars and interstellar matter in the galactic nucleus,
to be aligned with that of the orbit of the binary black holes, implying
that an accretion disc should generally be inclined to its binary's orbital
plane and should therefore precess.

It is still remarkable that OJ287 has a compact black hole binary of limited
life expectancy (Eq. 7) during the period (probably brief compared to the
age of the Universe) during which it is a luminous AGN.  This suggests that
the presence in a binary of the supermassive black hole may {\it cause} AGN
activity, rather than just being accidentally associated with it.  It is
possible to speculate that each black hole and its surrounding accretion
disc may facilitate accretion by the other black hole and disc.  The
gravitational or hydrodynamic mechanisms by which this may occur are surely
complex, but may have their origin in the fact that the gravitation of a
binary does not exert a central force, and does not conserve the angular
momentum of matter near it.

I thank NASA NAG 52682 and NSF AST 94-16904 for support and B. Margon for
discussions.
\vfil
\eject
\centerline{References}
\parindent=0pt
\def\ref{\medskip \hangindent=20pt \hangafter=1}
\ref
Bahcall, J. N., Dyson, F. J., Katz, J. I. \& Paczy\'nski, B. 1974 ApJL 189,
L17
\ref
Baykal, A., Anderson, S. F. \& Margon, B. 1993 AJ 106, 2359
\ref
Baykal, A., Boynton, P. E., Deeter, J. E. \& Scott, D. M. 1993 MNRAS 265,
347
\ref
Begelman, M. C., Blandford, R. D. \& Rees, M. J. 1980 Nature 287, 307 (BBR)
\ref
Katz, J. I. 1973 Nature PS 246, 87
\ref
Katz, J. I. 1980 ApJL 236, L127
\ref
Katz, J. I., Anderson, S. F., Margon, B. \& Grandi, S. A. 1982 ApJ 260, 780
(KAMG)
\ref
Levine, A. M. \& Jernigan, J. G. 1982 ApJ 262, 294
\ref
Sillanp\"a\"a, A., Haarala, S., Valtonen, M. J., Sundelius, B. \& Byrd, G.
G. 1988 ApJ 325, 628
\ref
Sillanp\"a\"a, A. {\it et al.} 1996 A\&A 305, L17
\vfil
\eject
\end
\bye